\documentstyle[11pt,paspconf,epsf]{article}

\def\x  {\times}

\newcommand{\ovec}[1]{{\mbox{\boldmath $#1$}}}

\newcommand{\Bvec}{\ovec{B}}
\newcommand{\uvec}{\ovec{u}}
\newcommand{\Bvecq}{\itol{\ovec{B}}}
\newcommand{\uvecq}{\ol{\ovec{u}}}
\newcommand{\Bvecs}{\ovec{B}'}
\newcommand{\uvecs}{\ovec{u}'}
\newcommand{\Escvec}{\mbox{\boldmath $\cal{E}$}}
\newcommand{\xvec}{\ovec{x}}
\newcommand{\rvec}{\ovec{r}}
\newcommand{\rvech}{{\mbox{\boldmath $\hat{r}$}}}
\newcommand{\bfnab}{{\mbox{\boldmath $\nabla$}}}

\newcommand{\stutz}[2]{\rule[#1]{0mm}{#2}}
\newcommand{\itover}[3]{\,\hspace{#3}#1{\!\hspace{-#3}#2}\stutz{0pt}{8.2pt}}
\newcommand{\ithat}[1]{\itover{\hat}{#1}{0.05em}}
\newcommand{\itol}[1]{\itover{\ol}{#1}{0.01em}}

\def\ol{\overline}

\def\qqq{\qquad\qquad\qquad}               
\def\q{\qquad}

\begin{document}

\title {Turbulent dynamo action in the high--conductivity limit:\\ a hidden dynamo}
\author{K.-H. R\"adler and U. Geppert\altaffilmark{1}}

\altaffiltext{1}{Astrophysikalisches Institut Potsdam, An der Sternwarte 16, 
14482 Potsdam, Germany}

\begin{abstract}
The paper deals with a simple spherical mean--field dynamo model of
$\alpha^2$--type in the case of high electrical conductivity.
A spherically symmetric distribution of turbulent motions is assumed
inside a spherical fluid body surrounded by free space, and some
complete form of the corresponding mean electromotive force is taken
into account. For a turbulence lacking reflectional symmetry finite growth rates  of magnetic fields 
prove to be possible even in the limit of perfect conductivity, that is,
the model corresponds to a fast dynamo. In accordance with the theorem by Bondi and Gold the dynamo--generated field is completely confined in the 
fluid body. 
\end{abstract}

\section{Introduction}

According to a finding by Bondi and Gold (1950)
the magnetic multipole moments which result from electric currents in 
a perfectly conducting fluid occupying a simply connected body 
cannot grow boundlessly. They remain for any fluid motion within bounds
determined by the initial magnetic flux through the surface of the body.
In a simple spherical mean--field dynamo model of $\alpha^2$--type proposed by Krause and Steenbeck (1967), however,
the magnetic field grows endlessly both inside and outside the fluid
body even in the limit of perfect conductivity. This conflict has been resolved in a more sophisticated model by R\"adler (1982) that
properly considers the specific structure of the mean electromotive force resulting from the constraints on the fluid motion at the boundary of the body. It was shown analytically that this structure
indeed ensures the boundedness of the magnetic multipole moments and
thus excludes an infinite growth of the magnetic field outside the 
fluid body in the high--conductivity limit. However, the question 
remained open whether or not a dynamo may work inside the fluid body, 
which then would have to be invisible, or hidden, in the sense that its
magnetic field is completely confined inside this body. Only a few arguments were given which support the conjecture that such a dynamo 
may exist.

In between the issue of the adjustment of a dynamo in the high--conductivity limit to the constraints posed by the Bondi and Gold theorem was also addressed in papers by Hollerbach, Galloway and 
Proctor (1995 and 1998).
They considered a dynamo in a spherical fluid shell with a chaotic flow 
surrounded by insulating space inside and outside and demonstrated that
it works also in the limit of high conductivity, in which then the
magnetic field in the insulating spaces vanishes. 

In the present paper we return to the mean--field model proposed by R\"adler (1982). After a few short explanations concerning the case of high conductivity and the mean--field concept in 
dynamo theory (Sections 2 and 3) and the simple model by Krause and Steenbeck 
(1967) (Section 4) we deliver a systematic treatment of our model and present numerical results giving 
evidence for the existence of a hidden fast dynamo in the above sense
(Section 5). Finally we discuss a few conclusions (Section 6).

\section{Dynamo action in the high--conductivity limit}

We consider here dynamo models consisting of an electrically conducting
fluid occupying a simply connected region surrounded by free space.
Let us assume that the magnetic flux density, $\Bvec $, is governed 
by the induction equation
\begin{equation}
\eta \bfnab^{2} \Bvec + \bfnab \x (\uvec \x \Bvec) 
 - \partial \Bvec / \partial t = \ovec{0} \, ,
 \q \bfnab \cdot \Bvec = 0 \, , 
\label{I1} \end{equation}
inside the fluid body, continues as a potential field
\begin{equation}
\Bvec = -\bfnab \phi \, , \q \Delta \phi = 0 \, ,
\label{I2} \end{equation}   
in outer space and vanishes at infinity. As usual, $\uvec$ means the velocity of the fluid and $\eta$ its magnetic diffusivity. 

For steady $\uvec$ we may expect solutions $\Bvec=\Re \big({\ithat{\Bvec}} \exp( pt )\big)$, 
where $\ithat{\Bvec}$ is a steady complex field and $p$ a complex quantity
independent of space and time coordinates. The real part $p_r$ of $p$ 
is the growth rate of $\Bvec$. We speak of a dynamo if there is
at least one solution $\Bvec$ with a non--negative $p_r$.

For many applications to cosmic objects the limit of high electrical
conductivity, that is $\eta \to 0$, deserves special interest. 
To define this limit more precisely, let us measure all length in units of $L$ being a typical scale of $\uvec$ or $\Bvec$, and the time in units
of $L/U$, with $U$ being a typical magnitude of $\uvec$. Then 
(\ref{I1}) applies with $\eta$ replaced by ${R_{\mbox{m}}}^{-1}$,
where $R_{\mbox{m}}$ is the magnetic Reynolds number defined by
\begin{equation}
R_{\mbox{m}} = U L / \eta \, , 
\label{I3} \end{equation}
and the high--conductivity limit corresponds to $R_{\mbox{m}} \to \infty$.

We mention two important aspects of dynamo action in the high--\-con\-duc\-ti\-vi\-ty 
limit. To explain the first one we compare dynamos which
differ only in the value of $R_{\mbox{m}}$. If then the largest growth rate
$p_{r}$ remains positive and takes a finite positive value as 
$R_{\mbox{m}} \to \infty$ we speak of a fast dynamo, otherwise, that is, 
if $p_{r}$ tends to zero or to a negative value, of a slow dynamo.

The second aspect is that in the case of perfect conductivity, 
${R_{\mbox{m}}}^{-1} = 0$, the magnetic field has to satisfy the requirements posed by the theorem by Bondi and Gold 
(1950). For the sake of simplicity we suppose 
the fluid body to be a sphere and formulate this theorem as proposed 
by R\"adler (1982). We use spherical coordinates
$r, \vartheta, \varphi$ and represent the potential $\phi$ introduced with (\ref{I2}) in the form
\begin{equation}
\phi = \sum \limits_{n = 1}^{\infty} \sum \limits_{m = -n}^{n} c^m_n
  r^{-(n+1)} Y^m_n(\vartheta, \varphi)
\label{I4} \end{equation}
with spherical harmonics $Y^m_n$ defined by
$Y^m_n(\vartheta, \varphi) = P^m_n(\cos \vartheta)
  \exp(\mbox{i} m \varphi)$
where the $P^m_n$ are associated Legendre polynomials. The $c^m_n$ 
are complex constants satisfying $c^{-m}_n = c^{m \ast}_n$ which define
multipole moments, e.g., those with $n = 1$ the dipole moment, with 
$n = 2$ the quadrupole moment etc. Depending on the motions at the 
boundary of the fluid body the $c^m_n$ may vary in time. 
According to the theorem by Bondi and Gold, however, they are bounded
in the sense that
\begin{equation}
|\, c^m_n \,| \le q^m_n \, , 
\label{I6} \end{equation}
with $q^m_n$ given by the initial distribution of the magnetic flux 
at the boundary. This in particular excludes any exponential growth
of the $c^m_n$.

A third aspect which concerns the conditions to be satisfied by $\Bvec$
at the boundary will be discussed below.

\section{The mean--field approach}

In cases in which the magnetic field and the fluid motion are of
turbulent nature, or show by other reasons complex structures in space or time, the mean--field approach to dynamo models has proved to be useful; 
see, e.g., Krause and R\"adler (1980). In this 
approach both the magnetic flux density $\Bvec$ and the fluid velocity $\uvec$ are considered as a sum of a mean field, $\Bvecq$ or $\uvecq$, defined by a proper averaging procedure, and a fluctuating part, $\Bvecs$ or $\uvecs$. Provided that the Reynolds averaging rules apply, we may conclude from equations (\ref{I1}) that
\begin{equation}
\eta \bfnab^{2} \Bvecq + \bfnab \x (\uvecq \x \Bvecq + \Escvec) 
 - \partial \Bvecq / \partial t = \ovec{0} \, , 
 \q \bfnab \cdot \Bvecq = 0 \, , 
\label{I10} \end{equation}
inside the fluid body, where $\Escvec$ is an electromotive force due to fluctuations,
\begin{equation}
\Escvec = \ol{\uvecs \x \Bvecs}\, .  
\label{I11} \end{equation}
For a given motion $\Escvec$ is a linear functional of $\Bvecq$. Under the usually accepted assumption of sufficiently weak variation of $\Bvecq$ in space and time it can be represented as
\begin{equation}
{\cal{E}}_i = a_{ij} \itol{B}_j 
  - b_{ijk} \partial \itol{B}_j / \partial x_k \, .
\label{I12} \end{equation}
Here we rely on Cartesian coordinates and use the summation convention. The tensors $a_{ij}$ and $b_{ijk}$ are, apart
from $\eta$, determined by $\uvecq$ and $\uvec'$. 

If, for instance, $\uvecq = \bf{0}$ and $\uvec'$ represents a homogeneous 
isotropic turbulence we have
\begin{equation}
\Escvec = \alpha \, \Bvecq - \beta \, \bfnab \x \Bvecq 
\label{I13} \end{equation}
with coefficients $\alpha$ and $\beta$ determined by $\uvecs$ or, what
is the same here, by $\uvec$ and being independent of space coordinates. The first term on the right--hand side
describes the $\alpha$--effect, that is, the occurrence of a mean 
electromotive force parallel or antiparallel to $\Bvec$ but vanishes if 
the turbulence is reflectionally symmetric. The second term gives rise 
to introduce a mean--field conductivity different from the usual one.  
Several results are available concerning the connection of $\alpha$ 
and $\beta$ with the properties of the $\uvec$--field. We mention in
particular those obtained in the second--order correlation approximation; see Krause and R\"adler (1980). 
In the high--conductivity limit, in this context defined by ${\eta{\tau_c}}/{\lambda_c^2} \to 0$ with ${\lambda_c}$ and ${\tau_c}$ being correlation length and time of the $\uvec$--field, $\alpha$ and $\beta$ then take the values $\alpha^{(0)}$
and $\beta^{(0)}$ given by
\begin{eqnarray}
\alpha^{(0)} &=&  \textstyle \frac{1}{3} \displaystyle \int_{0}^{\infty} 
 \ol{\uvec(\xvec,t) \cdot \big(\bfnab \x \uvec(\xvec,t-\tau)\big)} \,d \tau  
\nonumber \\
\label{I14} \\
\beta^{(0)} &=& \textstyle \frac{1}{3} \displaystyle \int_{0}^{\infty} 
 \ol{\uvec(\xvec,t) \cdot \uvec(\xvec,t-\tau)} \,d \tau \, .
\nonumber \end{eqnarray}
In general both $\alpha$ and $\beta$ do not vanish in this limit.

\section{A (too) simple model}

One of the simplest mean--field dynamo models, which can be treated 
analytically, has been proposed by Krause and Steenbeck 
(1967). The fluid body is supposed to be a sphere of
radius R, again surrounded by free space. Any mean motion of the fluid is ignored, ${\uvecq} = \bf{0}$, and thinking of homogeneous isotropic turbulence the electromotive force $\Escvec$ is taken in the form (\ref{I13}) with constant $\alpha$ and $\beta$. Specifying now 
equations (\ref{I10}) to this case we write simply $\Bvec$ instead of $\Bvecq$ and use $R$ and $R^2 /(\eta + \beta)$ as units of length and time. Looking then for solutions $\Bvec$ varying like $\exp (pt)$ with
$t$ we may reduce these equations to
\begin{equation}
\bfnab^2 \Bvec + C \bfnab \x \Bvec - p \Bvec = \ovec{0} \, , \q \bfnab \cdot \Bvec = 0\, ,
\label{I15} \end{equation}
where $C$ is a dimensionless measure of the $\alpha$--effect,
\begin{equation}
C = \alpha R /(\eta + \beta) \, .
\label{I16} \end{equation}
In this model equation (\ref{I15}) was completed by the requirements that $\Bvec$ is continuous across the boundary and has the structure given by (\ref{I2}) and (\ref{I4}) in outer space. The problem posed in this way has been investigated for axisymmetric $\Bvec$--fields by Krause and Steenbeck (1967) for the
steady case and by Voigtmann 
(1968) for the time--dependent case. A detailed treatment of the general case including non--axisymmetric time--dependent fields is given in Krause and R\"adler
(1980). We only mention here a few particular results. There are independent solutions $\Bvec$ which possess the form of single multipole fields, that is dipole fields, quadrupole fields etc. in outer space. In all cases $p$ is real, it takes a negative value for $C=0$ and increases monotonously with $|\, C\,|$, runs through zero for some marginal value of $|\, C\,|$ and behaves like $C^2$ as $C\to\infty$. Comparing all these solutions we find the smallest marginal value for a solution of dipole type. Denoting this 
value by $C_{\mathrm{crit}}$ we have 
\begin{equation}
C_{\mathrm{crit}} = 4.493 \, .
\label{I17} \end{equation}

Consider now the high--conductivity limit. Then the condition of
dynamo action resulting from (\ref{I16}) and (\ref{I17}) takes the form 
$|\,\alpha^{(0)}\,| R / \beta^{(0)} \ge 4.493$.
It seems well possible to satisfy this condition, that is, the dynamo 
may well work in this limit. The dimensional growth rate is then given by $\beta^{(0)} p / R^2$. Since in the case of a dynamo with growing magnetic fields $p$ must be positive, this growth rate is positive too,
that is, we have a fast dynamo.

The structure of the magnetic field depends, if its multipole
character is specified, only on the value of the parameter $C$ irrespective of that of $\eta$. The field is non--zero both inside and outside the fluid body, and this cannot change as $\eta \to 0$. If a dynamo works 
and the field grows exponentially in time then the $c^m_n$ introduced 
with (\ref{I4}) do so, too. This, however, is in conflict with the Bondi--Gold
theorem.

\section{The more sophisticated model}

\subsection{Basic equations}

One of the shortcomings of the model considered so far, which might be
the reason for this conflict, is the assumption of an electromotive 
force $\Escvec$ in the form (\ref{I13}) with constant $\alpha$ and 
$\beta$, which can be justified for homogeneous isotropic turbulence only. Near the boundary of the fluid body the turbulence must necessarily deviate 
from homogeneity and isotropy, and  $\Escvec$ has to take a more complex
form. As already mentioned, in a paper by R\"adler (1982) 
a modified model was studied with a spherically symmetric distribution 
of the turbulence. Then the radial direction necessarily occurs as preferred direction in the turbulence and therefore we have a  more complex form of $\Escvec$.
It was shown that with this modification the conflict is resolved. More precisely, with these assumptions any growth of the magnetic field at the boundary of the fluid body or in outer space can be ruled out. 

Turning now to this model we assume again that the mean magnetic flux 
density $\Bvecq$ is governed by equations (\ref{I10}) in a spherical
fluid body $r<R$ and continues as a potential field as given by 
(\ref{I2}), or (\ref{I4}), in the outer space $r>R$. We further exclude 
any mean motion, $\uvecq = \bf{0}$, and specify the electromotive force 
$\Escvec$ in accordance with a spherically symmetric turbulence. For 
the sake of simplicity we write in the following again $\Bvec$ instead
of $\Bvecq$, and $\uvec$ instead of $\uvecs$, and we rely again on 
spherical coordinates $r, \vartheta$, $\varphi$.

Spherical symmetry of the turbulence is understood here in the sense 
that all averaged quantities determined by the $\uvec$--field are 
invariant under arbitrary rotations of this field about arbitrary axes 
through the center $r=0$ of the fluid body. From this definition we may 
conclude by standard reasoning (see, e.g., Krause and R\"adler 
(1980)) that the electromotive force $\Escvec$ must have the 
form
\begin{eqnarray}
\Escvec &=& - \,\alpha_1 \Bvec 
    - \alpha_2(\rvech \cdot \Bvec)\rvech 
    - \gamma \rvech \x \Bvec  
\nonumber \\
    & & - \,\beta_1 \bfnab \x \Bvec 
    - \beta_2 \big(\rvech \cdot(\bfnab \x \Bvec)\big) \rvech 
    - \delta \rvech \x (\bfnab \x \Bvec)   
\label{M1} \\
    & & - \,\beta_1^r \big( \rvech \cdot (\nabla \Bvec)^s \big)
    - \beta_2^r 
      \big( \rvech \cdot \big( \rvech \cdot (\nabla \Bvec)^s \big)\big) \rvech  
    - \delta^r 
      \rvech \x \big( \rvech \cdot (\nabla \Bvec)^s \big)
\nonumber     
\end{eqnarray}         
with scalar coefficients ${\alpha_1}, {\alpha_2}, ..., {\delta^r}$ 
determined by $\uvec$ and depending on position only through $r$ but
not through $\vartheta$ or $\varphi$. Here $\rvech$ is the radial unit 
vector and ${({\nabla}{\Bvec})^s}$ the symmetric part of the gradient tensor of ${\Bvec}$ so that, if we refer to Cartesian coordinates, 
${\big(\rvech \cdot {(\nabla \Bvec)}^s\big)}_i = \frac{1}{2} 
 {\hat{r}}_j (\partial B_i/ \partial x_j + \partial B_j/ \partial x_i)$.
Some more details concerning the derivation of (\ref{M1}) are given in R\"adler (1982). If the turbulence is reflectionally
symmetric about planes containing the center of the body 
$\alpha_1, \alpha_2, \delta, \beta_1^r$ and $\beta_2^r$ are equal to zero.
In the case of homogeneous isotropic turbulence only the coefficients $\alpha_1$ and $\beta_1$ can be non--zero and the others have to vanish, so that we return to (\ref{I13}).
 
Later we will also assume that the turbulence is steady, that is, that all averaged quantities depending on $\uvec$ are invariant under shifts along the time axis. In this case the coefficients 
$\alpha_1, \alpha_2, ..., \delta^r$ are independent of time.

Even if we accept again the second--order correlation approximation the
calculation of the coefficients $\alpha_1, \alpha_2, ..., \delta^r$ for points near the boundary of the fluid body is, at least for finite conductivity, rather complex. We note here only results for the high--conductivity limit, defined as above by 
$\eta \tau_c /\lambda_c^2 \to 0$, which were already obtained in a slightly different form by R\"adler (1982). Denoting the
mentioned coefficients in this limit by $\alpha_1^{(0)}, \alpha_2^{(0)},
 ..., \delta^{r(0)}$  we have  
\begin{eqnarray}
\alpha_1^{(0)} &=& {\textstyle\frac{1}{2}} ( a_\parallel + {\tilde a}_\parallel 
  - \frac{d}{r} ) \, , \q
 \alpha_2^{(0)} = {\textstyle \frac{1}{2}} ( 4 a_\perp - a_\parallel 
  - 3 {\tilde a}_\parallel - 3 \frac{d}{r} ) \, ,
\nonumber \\
\gamma^{(0)} &=& - \frac{b_\perp - b_\parallel}{r} 
  + {\textstyle\frac{1}{2}} \frac{d b_\parallel}{d r} 
  + {\textstyle\frac{1}{2}} c \, ,
\nonumber \\
\beta_1^{(0)} &=& {\textstyle\frac{1}{2}} ( b_\perp + b_\parallel ) \, , \q
  \beta_2^{(0)} = - {\textstyle\frac{1}{2}} \delta^{r(0)}
  = {\textstyle\frac{1}{2}} ( b_\perp - b_\parallel ) \, ,
\label{M2} \\
\delta^{(0)} &=& {\textstyle\frac{1}{2}} \beta_1^{r(0)} = - {\textstyle\frac{1}{4}} d \, , \q
  \beta_2^{r(0)} = 0 \, 
\nonumber \end{eqnarray}
and
\begin{eqnarray}
a_\parallel &=&\int_{0}^{\infty} 
  \ol{\uvec_\parallel (\xvec, t) 
  \cdot {\big(\bfnab \x \uvec(\xvec, t-\tau)\big)}_\parallel } \,d \tau \, ,
\nonumber \\ 
{\tilde a}_\parallel &=& \int_{0}^{\infty} 
  \ol{\uvec_\parallel (\xvec, t-\tau) 
  \cdot {\big(\bfnab \x \uvec(\xvec, t)\big)}_\parallel } \,d \tau  \, , 
\nonumber \\
a_\perp &=& \textstyle\frac{1}{2} \displaystyle \int_{0}^{\infty} 
  \ol{\uvec_\perp (\xvec, t) \cdot 
  {\big(\bfnab \x \uvec(\xvec, t-\tau)\big)}_\perp } \,d \tau \, ,
\nonumber \\
b_\parallel &=& \int_{0}^{\infty} 
  \ol{\uvec_\parallel (\xvec, t) 
  \cdot \uvec_\parallel (\xvec, t-\tau) } \,d \tau \, ,
\hspace{-8em}\label{M3} \\
b_\perp &=& \textstyle\frac{1}{2} \displaystyle \int_{0}^{\infty} 
  \ol{\uvec_\perp (\xvec, t) 
  \cdot \uvec_\perp (\xvec, t-\tau) } \,d \tau \, ,
\nonumber \\
\, c &=& \int_{0}^{\infty}\ 
  \Big(\, \ol{\big(\rvech \cdot \uvec (\xvec, t)\big)
  \big(\bfnab \cdot \uvec (\xvec, t-\tau)\big) }  
   - \ol{\big(\rvech \cdot \uvec (\xvec, t-\tau)\big)
   \big(\bfnab \cdot \uvec (\xvec, t)\big)} \,\Big) \,d \tau \, ,
\nonumber \\
\, d &=& \int_{0}^{\infty}
  \rvech \cdot \ol{\big(\uvec (\xvec, t) \x \uvec (\xvec, t-\tau)\big)} \,d \tau \, ,
\nonumber \end{eqnarray}
with $\uvec_\parallel = ( \rvech \cdot \uvec ) \rvech$ 
and $\uvec_\perp = \uvec - \uvec_\parallel$ and analogous definitions
of ${(\bfnab \x \uvec)}_\parallel$ and ${(\bfnab \x \uvec)}_\perp$. 
According to our assumptions the $\alpha_1^{(0)}, \alpha_2^{(0)}, ..., \delta^{r(0)}$, like the averaged quantities under the integrals, depend on $\xvec$ via $r$ only. 

It seems natural to assume that $b_\parallel$ and $b_\perp$ are non--negative everywhere.
At the boundary of the fluid body we have
$\rvech \cdot \uvec = 0$ and therefore 
\begin{equation}
a_\parallel = {\tilde a}_\parallel = b_\parallel
  = d b_\parallel / d r = c = 0  \q \mbox{at} \q r = 1 \, . 
\label{M4} \end{equation}

\subsection{Reduction of the basic equations}

Let us now represent the magnetic flux density ${\Bvec}$ as a sum of a poloidal and toroidal part,
\begin{equation}
\Bvec = -\bfnab \x (\rvec \x \bfnab S)- (\rvec / R) \x \bfnab T \, ,
\label{R1} \end{equation}
where $\rvec = r \rvech$, and expand the defining scalars $S$ and $T$
in series of spherical harmonics $Y^m_n(\vartheta, \varphi)$. 
It can easily be followed up that the equations and conditions governing
$\Bvec$ imply no coupling between contributions to $\Bvec$ differing 
in $n$ or $m$ so that we may restrict ourselves to the simple solutions
defined by 
\begin{equation}
S = S(r,t) Y^m_n (\vartheta, \varphi) \, , \; 
  T = T(r,t) Y^m_n (\vartheta, \varphi) \, , \;
  n \ge 1 \, , \; |\, m \,| \le n \, .
\label{R2} \end{equation}

Due to the factor $R$ in (\ref{R1}) the dimensions of $S$ and $T$
coincide. Preparing the definition of dimensionless quantities we 
introduce the constants $\alpha^0$ and $\beta^0$ with the dimension of 
a velocity and a magnetic diffusivity. We will assume homogeneous 
isotropic turbulence in some central region of the fluid body and 
identify these constants with $\alpha^{(0)}$ and $\beta^{(0)}$ given by (\ref{I14})
for this region. In the following we measure all lengths in units of
$R$, the time in units of $R^2 / \beta^0$, further the 
$\alpha_1, \alpha_2, a_\parallel, {\tilde a}_\parallel, a_\perp$ 
in units of $\alpha^0$, 
the $\delta, \beta_1^r, \beta_2^r, d$ in units of $\alpha^0 R$, 
the $\beta_1, \beta_2, \delta^r, b_\parallel, b_\perp$ in units of $\beta^0$, and $\gamma, c$ in units of $\beta^0 / R$.      

Using standard methods we may then reduce equations (\ref{I10}) for $\Bvec$ to
\begin{eqnarray}
\varepsilon D S + U_S + C U_T - \partial S / \partial t &=& 0 
\nonumber \\
& & \qqq \mbox{for} \q r < 1
\label{R3} \\
\varepsilon D T + C V_S + V_T - \partial T / \partial t &=& 0  
\nonumber \end{eqnarray}
with
\begin{equation}
\varepsilon = \eta / \beta^0 \, , \q
  C = \alpha^0 R / \beta^0\, , 
\label{R4} \end{equation}
\begin{equation} 
D f = \frac{1}{r} \frac{\partial^2 ( r f )}{\partial r^2}  
  - \frac{n(n+1)}{r^2} \, f  
\label{R5} \end{equation}
and
\begin{eqnarray}
U_S &=& \frac{\gamma}{r} \frac{\partial (rS)}{\partial r}
  + \beta_1 D S
  - \frac{\delta^r}{2 r^2} \big( 2 S - r^2 \frac{\partial^2 S}{\partial r^2}
    - n(n+1) \, S \big) \, ,\nonumber\\
   &\phantom{=}&\label{R6a}\\
U_T &=& - \,\alpha_1 T 
  + \frac{\delta}{r} \frac{\partial (rT)}{\partial r}
  + \frac{\beta_1^r}{2 r} \big( 2 T - \frac{\partial (rT)}{\partial r} \big) \, ,
  \nonumber
\end{eqnarray}

\begin{eqnarray}
V_S &=& \frac{1}{r} \frac{\partial}{\partial r} 
   \left( \alpha_1 \frac{\partial (rS)}{\partial r} \right) 
  - (\alpha_1 + \alpha_2) \frac{n(n+1)}{r^2} \, S
  - \frac{1}{r} \frac{\partial}{\partial r}
   (\delta r D S ) 
\nonumber\\
   &\phantom{=}&- \,\frac{1}{2r} \frac{\partial}{\partial r} 
   \left(\frac{\beta_1^r}{r} 
   \big(2 S - r^2 \frac{\partial^2 S}{\partial r^2} - n(n+1) \, S \big) \right)
\nonumber\\
   &\phantom{=}&- \,( \beta_1^r + \beta_2^r )
   \frac{n(n+1)}{r} \frac{\partial}{\partial r} (\frac{S}{r}) \, , 
\nonumber\\
   &\phantom{=}&\label{R6b}\\
V_T &=& \frac{1}{r} \frac{\partial ( \gamma r T )}{\partial r}
  + \frac{1}{r} \frac{\partial}{\partial r}
   \big( \beta_1 \frac{\partial ( r T )}{\partial r}\big)
\nonumber\\
  &\phantom{=}&- \,( \beta_1 + \beta_2 ) \frac{n(n+1)}{r^2} \, T 
               -\frac{1}{2 r} \frac{\partial}{\partial r} 
  \big( \delta^r ( 2 T - \frac{\partial ( r T )}{\partial r} ) \big)\, . 
\nonumber
\end{eqnarray}
Equations (\ref{I2}) for the outer space are equivalent with
\begin{equation}
S = S(r = 1) / r^{n+1} \, , \q T = 0 \q \mbox{for} \q r > 1 \, . 
\label{R7} 
\end{equation}
Note that $S(r = 1)$ coincides with $c^m_n / R^{n+1}$ for the chosen 
$n$ and $m$. 

As a consequence of the divergence relation for $\Bvec$ its normal 
component and thus $S$ has to be continuous across the boundary $r = 1$. For finite conductivity of the fluid, that is $\, \varepsilon > 0$, 
we may exclude surface currents so that the tangential components have 
to be continuous too, and so $\partial S/ \partial r$ and $T$. Together with (\ref{R7}) we may conclude that
\begin{equation}
\partial S / \partial r + (n+1) S = T = 0 \q  
  \mbox{at} \q r = 1 \, . 
\label{R8} \end{equation}
The original problem of the determination of $\Bvec$ occurs then
as the problem of solving the equations (\ref{R3}), completed by (\ref{R6a}) and  (\ref{R6b}), with the boundary conditions (\ref{R8}). 

In the case of perfect conductivity, $\varepsilon = 0$, again the 
continuity of $S$ has to be required. However, surface currents can
no longer be excluded, which correspond to discontinuities of 
$\partial S/ \partial r$ and $T$. More precisely, such currents are
given by 
$(1 / \mu R^2) ([\partial S/ \partial r] \rvech \x \bfnab Y_n^m 
+ [T] \bfnab Y_n^m $),
where $\mu$ means the magnetic permeability of the fluid and $[f]$ stands for $f(r = 1 + 0) - f(r = 1 - 0)$.
In this case the equations (\ref{R3}) have to be completed by conditions 
which we will formulate later.

\subsection{Compatibility with the Bondi--Gold theorem}
 
Remaining with the case of infinite conductivity we first demonstrate,
repeating the ideas described by R\"adler (1982), that our 
model is compatible with the Bondi--Gold theorem. For this purpose 
we consider the first equation (\ref{R3}), completed by 
(\ref{R6a}), for $r \to 1$. Putting there $\varepsilon = 0$, 
inserting $\alpha_1, \alpha_2, ...,\delta^r$ according to (\ref{M2}) and using (\ref{M4}) we find
\begin{equation}
\partial S / \partial t = -\, n(n+1)\,b_\perp S \q 
  \mbox{at} \q  r = 1 \, , 
\label{R9} \end{equation}
and hence
\begin{equation}
S( r = 1, t ) = S( r = 1 , 0 )\, 
 \exp \big( - \textstyle\frac{1}{2} n(n+1)\,b_\perp (r = 1) \, t \, \big)\, .
\label{R10} \end{equation}
Remembering that $S( r = 1 ) = c_n^m / R^{n+1}$  and that $b_\perp $ is non--negative 
we see that (\ref{I6}) is indeed satisfied.

\subsection{Further specification of the model} 

We now turn our attention to the question whether in the limit of
infinite conductivity magnetic fields can grow inside the fluid body. 
We rely on equations (\ref{R3}) to (\ref{R7}), put $\varepsilon = 0$
and insert $\alpha_1, \alpha_2, \cdots, \delta^r$ according to (\ref{M2}). With the idea to have a model with a minimum of free
parameters we choose simply
\begin{equation}
a_\parallel = {\tilde a}_\parallel = b_\parallel = f \, , \q
  a_\perp = b_\perp = 1 \, , \q
  c = d = 0
\label{R12}\end{equation}
with
\begin{equation}
 f = \left\{\,
\begin{array}{lll}
1\, ,                           & \quad\mbox{for}\quad &  0 \leq r \leq 1 - q \\
10\xi^3 - 15\xi^4 + 6\xi^5  \, , \quad \xi = (1 - r)/q\, , & \quad\mbox{for}\quad &  1 - q \leq r \leq 1\; .
\end{array}
\right.
\label{R13}\end{equation}
This corresponds to the assumption of homogeneous isotropic turbulence
inside the spherical region $0 \leq r \leq 1 - q$ and of inhomogeneous 
anisotropic turbulence in the surrounding shell $1 - q \leq r \leq 1$.
Note that $f = df / dr = d^2f / dr^2 = 0$ at $r = 1$ and therefore
(\ref{M4}) is satisfied.
Finally we assume that there is initially no magnetic field at the boundary of the fluid body, that is $S = T = 0$ at $r = 1$. Due to (\ref{R9}) we have then  $S = 0$ at $r = 1$ at any time.
Starting from the second equation (\ref{R3}) together with (\ref{R6b}) and using 
(\ref{M2}), (\ref{R12}) and (\ref{R13}) we arrive at a relation for  $ T$ analogous 
to (\ref{R9}) from which we may conclude that also $T = 0$ at $r = 1$ at any time.
The behaviour of $\Bvec$ is then governed by (\ref{R3}), completed by
(\ref{R6a}), (\ref{R6b}), (\ref{M2}), (\ref{R12}) and (\ref{R13}), and the role of boundary conditions is taken by $S = T = 0$ at $r = 1$. (We note that the remarks concerning the boundary condition for $S$ in 
R{\"a}dler (1982), Section 3.6, are incorrect.)

\subsection{Numerical results}

\begin{figure}[h]
\plotfiddle{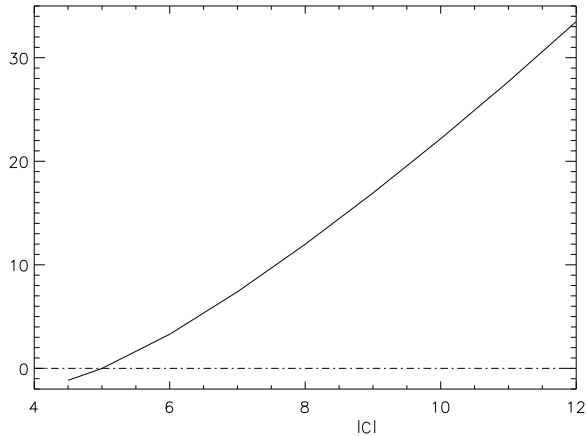}{6cm}{0}{50}{50}{-134}{0}
\caption{The growth rate $p$ versus $|\, C \,|$}
\label{F1}
\end{figure}

The problem posed in this way has been investigated numerically. There are indeed solutions $\Bvec$, or $S$ and $T$, which for sufficiently 
large $|\, C \,|$ grow exponentially in time. The growth rate $p$ introduced as above again proves to be real. The fastest--growing solution for a given $|\, C \,|$ is always one of dipole type, that is $n = 1$. We restrict our attention in this paper to such solutions and, furthermore, to the case $q = 0.2$.
The dependence of the growth rate $p$ on $|\, C \,|$ is shown in 
Figure \ref{F1}. Clearly dynamo action is possible if 
$|\, C \,| \geq C_{\mathrm{crit}}$ with
\begin{equation}
C_{\mathrm{crit}} = 5.002 \, .
\label{R14}\end{equation}

\begin{figure}[b]
\mbox{
\hspace{-.165\textwidth}
\epsfxsize=.78\textwidth \epsfbox{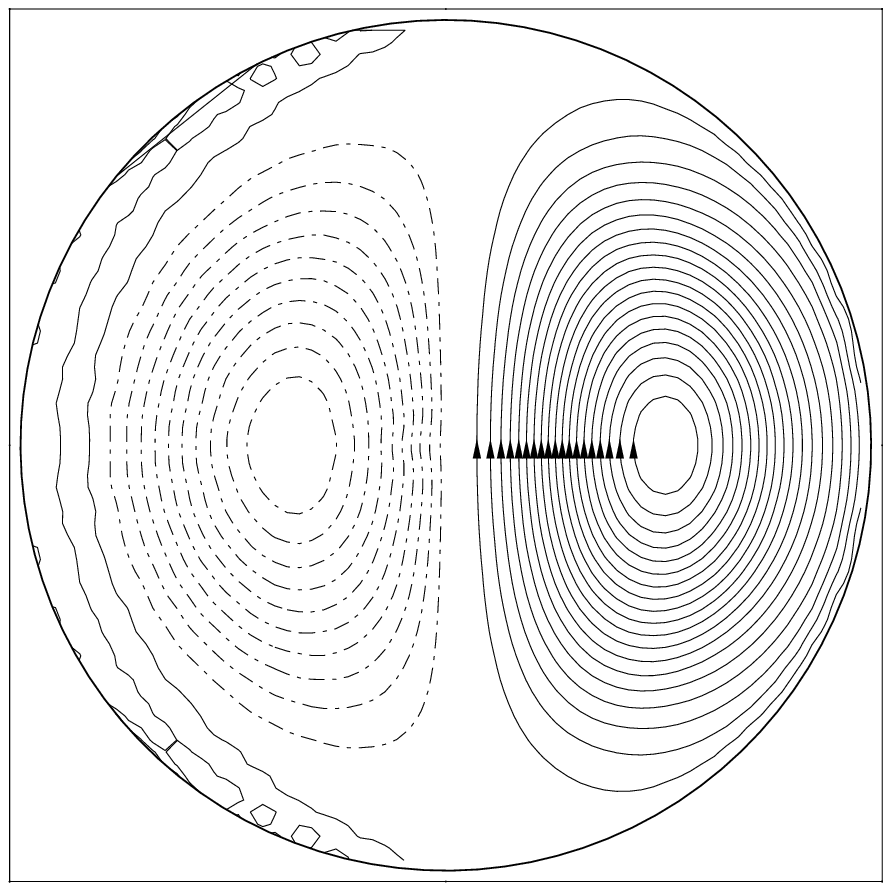}
\hspace{-.275\textwidth}
\raisebox{.003\textwidth}{\epsfxsize=.775\textwidth \epsfbox{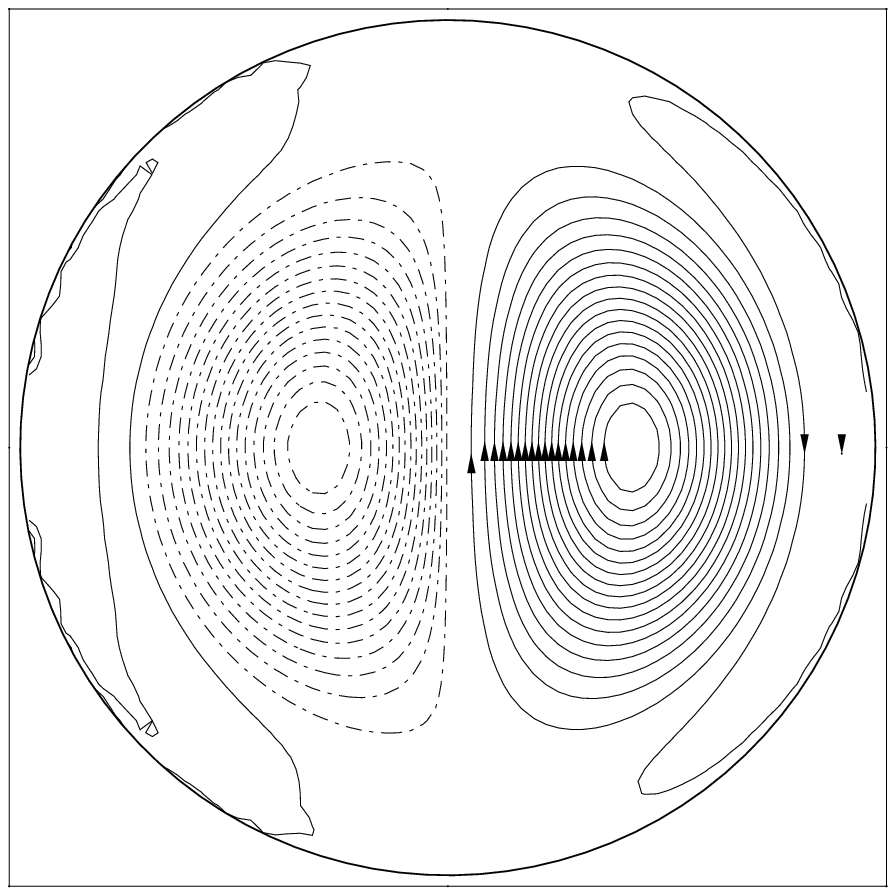}}
}
\caption{Field lines of the poloidal part (right) and isolines of the toroidal
part (left) of the magnetic field. The solid isolines correspond to positive,
the broken ones to negative values of the $\varphi$--component of the field.
Left panel $C = 8$, right panel $C = 12$. }
\label{F3}
\end{figure}

Let us define an effective radius $R_{\mathrm{eff}}$ of our model by equating
of $\alpha^0 R_{\mathrm{eff}} / \beta^0$ to the value of $C_{\mathrm{crit}}$ 
for the simple model discussed above given by (\ref{I17}), that
\begin{figure}
\mbox{
\hspace{-.073\textwidth}
\epsfxsize=.56\textwidth \epsfbox{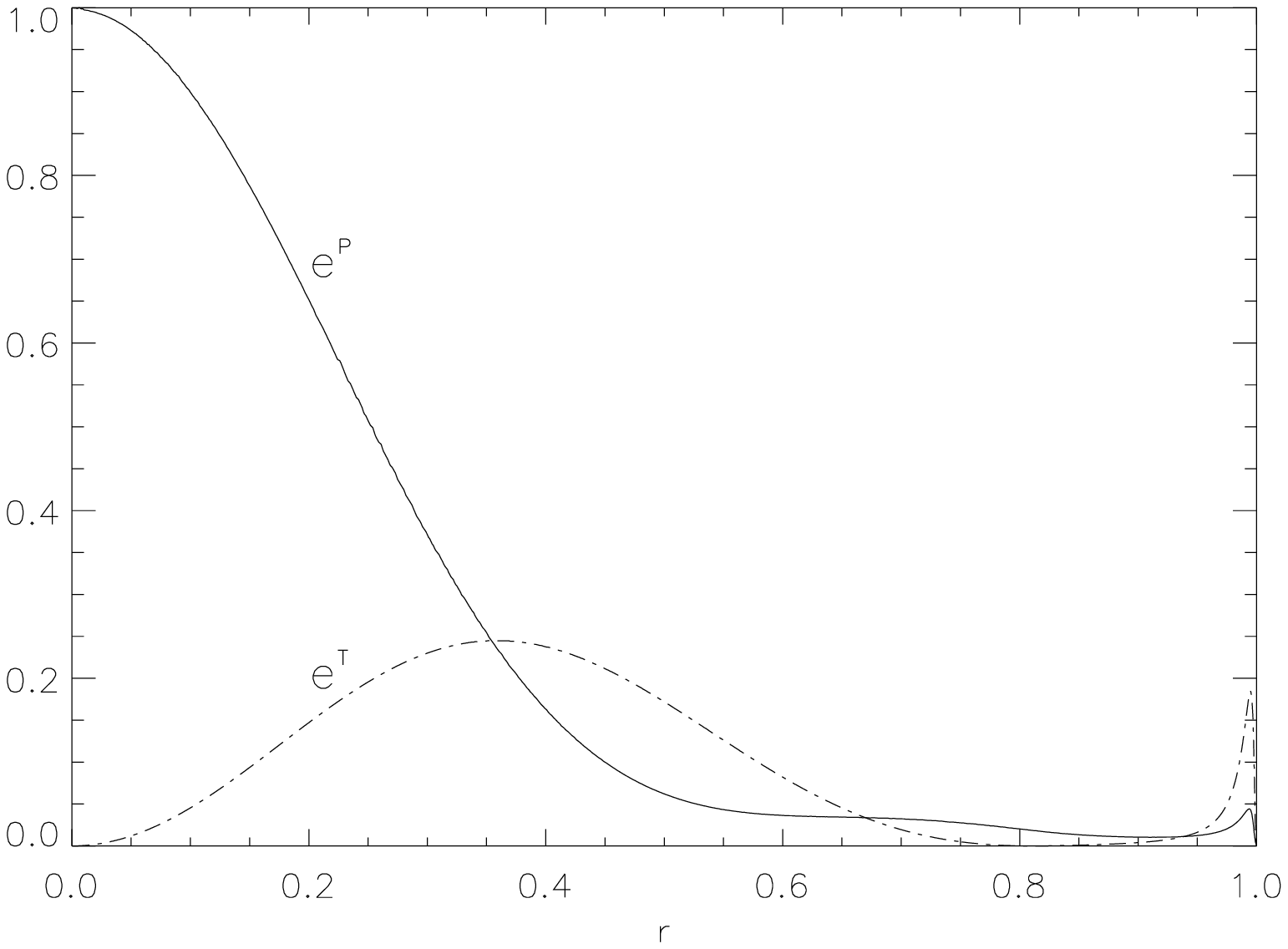}
\hspace{-.06\textwidth}
\epsfxsize=.56\textwidth \epsfbox{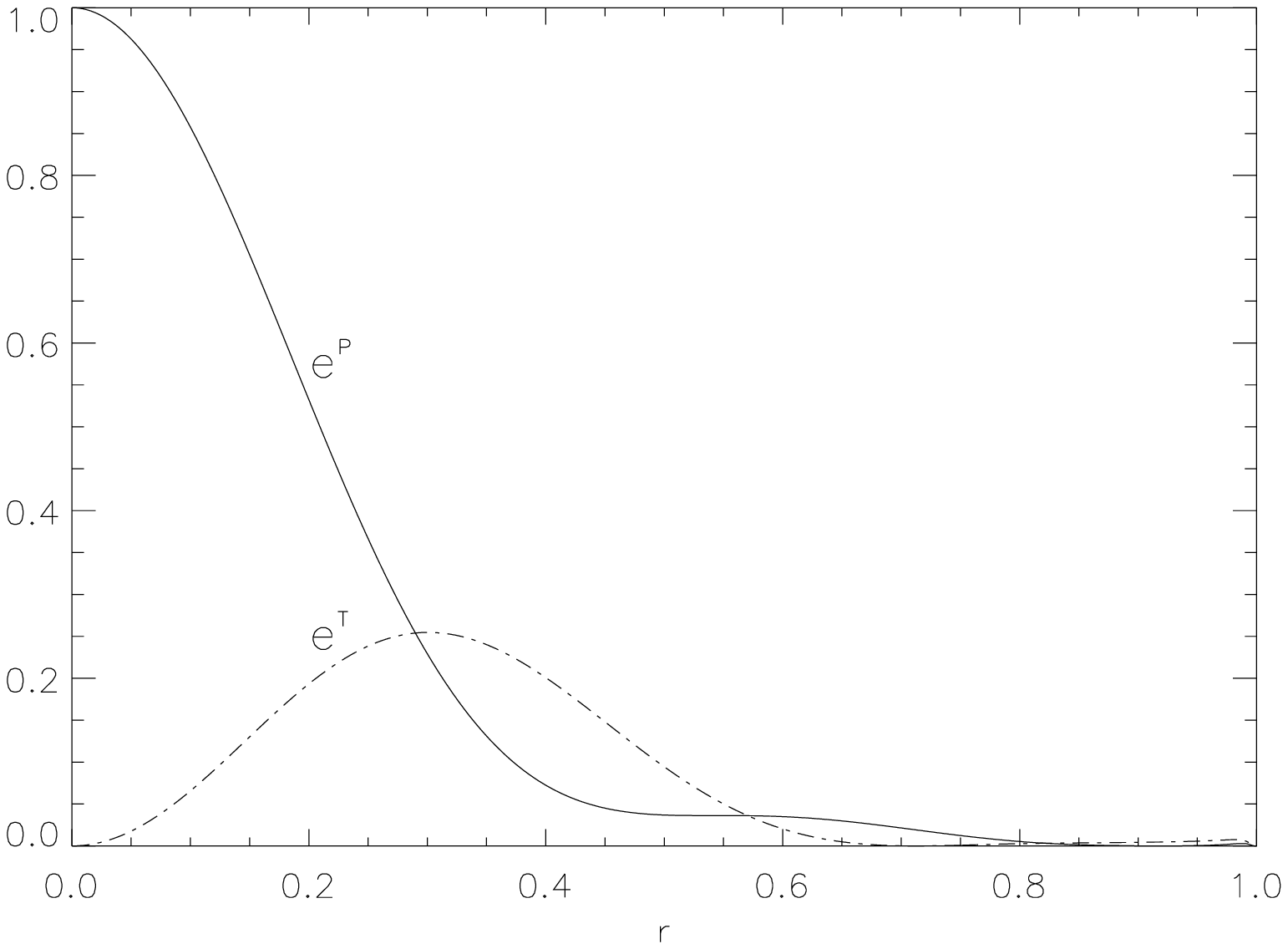}
}
\caption{The energy densities $e^{\mathrm{P}}$ and $e^{\mathrm{T}}$ of the 
poloidal and toroidal parts of the magnetic field in arbitrary units
in dependence of the fractional radius $r$. Left panel $C = 8$, right panel $C = 12$.}
\label{F2}
\end{figure}
is $\alpha^0 R_{\mathrm{eff}} / \beta^0 = 4.493$. Comparing this with the 
corresponding relation for the modified model under consideration, that is
$\alpha^0 R / \beta^0 = 5.002$, we find $R_{\mathrm{eff}} / R = 0.90$.
This means that the shell with the fractional radius between $0.8$
and $1$ showing deviations from a homogeneous isotropic turbulence is less effective for dynamo action than the central region of the fluid body without such deviations.
Figures \ref{F2} and \ref{F3} show the radial distribution of the energy densities of the poloidal and the toroidal parts of the magnetic field and field pictures 
for two examples with $C > C_{\mathrm{crit}}$.

\pagebreak
It would be interesting to study a sequence of dynamo models with
positive $\varepsilon$ approaching zero. In this context we must have 
in mind that $\varepsilon$ occurs not only with the dissipation terms $\varepsilon D S$ and $\varepsilon D T$ in (\ref{R3}). In addition
the coefficients $\alpha_1, \alpha_2, \cdots, \delta_1^r$
entering $U_S, U_T, V_S$ and $V_T$ via (\ref{R6a}) and (\ref{R6b}) depend on the 
magnetic diffusivity, that is, on a parameter proportional to
$\varepsilon$. There are, however no results available which describe  this dependence in the neighbourhood of a jump of the magnetic diffusivity as it occurs at the boundary.

Several numerical calculations have been carried out on the basis of 
equations (\ref{R3}) including the dissipation terms but using
expressions for 
$\alpha_1, \alpha_2, \cdots, \delta_1^r$ as given by (\ref{M2}) and (\ref{M3}), that is, 
for infinite conductivity. Of course, in these calculations the boundary conditions (\ref{R8}) have been used. Indeed the solutions obtained in this way approach the corresponding ones with $\varepsilon = 0$ in a very satisfying manner as $\varepsilon \to 0$. The only difference occurs in the distribution of the electric currents near the boundary, what is understandable since for $\varepsilon = 0$ a part of them occurs as surface currents.

It is interesting to observe how the magnetic energy concentrates
itself more and more inside the conducting body as $\varepsilon \to 0$.
Table \ref{T1} shows for one example the dependence of the ratio 
of the energies $E_{\mathrm{out}}$ and $E_{\mathrm{tot}}$ in the outer and in all space on $\varepsilon$.
\begin{table}
\begin{center} 
\begin{tabular}{c@{\hspace{6mm}}c}
\tableline\\*[-2mm]
$\varepsilon\; $ &  $\; E_{\mathrm{out}}/E_{\mathrm{tot}} $    \\*[2mm]
\tableline     \\*[-1mm]
  $10^{-2}\; $  &  $\; 8.6 \cdot 10^{-6} $       \\
  $10^{-3}\; $  &  $\; 4.9 \cdot 10^{-7} $        \\
  $10^{-4}\; $  &  $\; 4.8 \cdot 10^{-9} $        \\*[2mm]
\tableline
\tableline      
\end{tabular}
\end{center} 
\caption{The energy ratio $E_{\mathrm{out}} / E_{\mathrm{tot}}$ in dependence on $ \varepsilon $ for $ C = 12 $. }
\label{T1}
\end{table}

\section{Conclusions}
As already shown in an earlier paper by R\"adler (1982), the mean--field model reconsidered here, in contrast to a more simplified one, meets the requirements posed by the 
Bondi--Gold theorem. It was, however, only conjectured but not really 
demonstrated that it admits dynamo action inside the fluid body in the 
high--conductivity limit. The present paper clearly demonstrates the
possibility of a fast dynamo inside the fluid body, whose magnetic field is then, as required by the Bondi--Gold theorem, completely 
confined in the fluid body and in that sense invisible from outside.    

The model is also in another respect, which is not necessarily connected
with the high--conductivity limit, of interest for the fundamentals of
mean--field dynamo theory. In most of the mean--field dynamo models elaborated
in view of cosmic objects only a few contributions to the
mean electromotive force have been taken into account, and others were  cancelled in the vague hope that they are of minor importance. 
We demonstrate here the possibility of dynamo action in an idealised model 
which is consistent in the sense that it includes all contributions to the
electromotive force which occur under this idealisation.

It remains to be discussed whether the concentration of the 
dynamo--ge\-ne\-ra\-ted magnetic field in the interior of a highly conducting
body, which was demonstrated above, may indeed occur in
astrophysical bodies. 

\bigskip
\bigskip
\bigskip
{\bf Acknowledgment}\\
The authors are indebted to Dr. M. Rheinhardt  for helpful discussion and for technical assistance.



\end{document}